\documentclass[aps,prl,twocolumn,showpacs]{revtex4}
\usepackage{amsbsy,latexsym}
\usepackage{amsfonts}
\usepackage{amssymb}
\usepackage[mathscr]{eucal}

\newcommand{\al}{\alpha}

\newcommand{\la}{\lambda}
\newcommand{\id}{\mathbb I}

\newcommand{\tr}{{\rm tr}\,}

\newcommand{\ket}[1]{\vert #1 \rangle}
\newcommand{\bra}[1]{\langle #1 \vert}

\newcommand{\D}[3]{\mathfrak D^{(#1)}_{#2 #3}}

\begin{document}
\title{Quantum reverse-engineering and reference frame alignment without non-local correlations}

\author{E.~Bagan}
 %\email{bagan@ifae.es}
 \affiliation{Grup de F{\'\i}sica Te{\`o}rica \& IFAE,
Facultat de Ci{\`e}ncies, Edifici Cn, Universitat Aut{\`o}noma de
Barcelona, 08193 Bellaterra (Barcelona) Spain}
\author{M.~Baig}
 %\email{baig@ifae.es}
 \affiliation{Grup de F{\'\i}sica Te{\`o}rica \& IFAE,
Facultat de Ci{\`e}ncies, Edifici Cn, Universitat Aut{\`o}noma de
Barcelona, 08193 Bellaterra (Barcelona) Spain}

\author{R.~Mu{\~n}oz-Tapia}
 %\email{rmt@ifae.es}
\affiliation{Grup de F{\'\i}sica Te{\`o}rica \& IFAE, Facultat de Ci{\`e}ncies,
Edifici Cn, Universitat Aut{\`o}noma de Barcelona, 08193 Bellaterra
(Barcelona) Spain}
%\date{\today}

\begin{abstract}
Estimation of unknown qubit elementary gates and  alignment of
reference frames are formally the same problem. Using quantum
states made out of $N$ qubits, we show that the theoretical
precision limit for both problems, which behaves as $1/N^{2}$, can
be asymptotically attained  with a covariant protocol that
exploits the quantum correlation of internal degrees of freedom
instead of the more fragile entanglement between distant parties.
This cuts by half the number of qubits needed to achieve the
precision of the dense covariant coding protocol.
\end{abstract}

\maketitle

Quantum resources are scarce goods and, as such, one has to make
sure they are used in the most efficient way. Optimal management
of systems and resources in quantum communication and state
estimation is, hence, a must. This subject has been addressed
extensively in the literature but only recently for the specific
type of problems that we will deal with: estimation of unitary
transformations on qubits. Given an unknown one-qubit gate (a
black box) which we may apply $N$ times over a number of qubits,
we are confronted with the reverse-engineering problem of  finding
out the hidden $SU(2)$ transformation performed by the gate. It
was shown in~\cite{ajv} that the optimal estimation is attained by
acting on a suitable maximally entangled state of $2N$ qubits (see
Eq.~\ref{maxentangled} below) and performing a collective
measurement on the $2N$ qubits. Note that in this protocol half of
the qubits are left untouched before the final measurement.

Closely linked to this reverse-engineering issue is the problem of
transmitting data that cannot be digitalized. This arises, for
instance, when someone (Alice) attempts to transmit the direction
of an arrow  to a distant party (Bob) with whom there is no shared
reference frame~\cite{direction}. In this situation, the
transmission of the information is only possible if the quantum
carrier is itself an arrow of some sort (e.g.,~an electron, which
has spin and magnetic dipole moment pointing along a specific
direction). A generalization of this problem consists of
transmitting the orientation of three orthogonal axes, i.e., a
trihedron, which we may view as a spatial reference frame
(throughout this paper we will use the word trihedron for
brevity). This problem is easily seen to be formally equivalent to
that of estimating a $SU(2)$ transformation. This is not at all
surprising because the group of proper rotations and $SU(2)$ are
locally isomorphic. It is important to realize that, likewise the
arrow example, the carrier of the information must be a quantum
system with intrinsic orientation (e.g.,~a hydrogen atom or a
system of electrons in a sufficiently asymmetric angular momentum
state), since Alice and Bob are assumed not to share any reference
frame.

This problem was first tackled in~\cite{frames-ps,frames-bbm} for
an atom or a system of $N$ spins under the simplification
assumption that the set of all the allowed signal states spans each $SU(2)$
irreducible representation exactly once. The transmission error of
the best covariant protocol was shown to vanish as $1/N$. This is
a somewhat puzzling result, because one can easily devise
non-covariant protocols that perform much better~\cite{frames-ps2}
(the corresponding error vanishes as $1/N^2$), despite of the fact
that covariance and optimality are generally regarded as
compatible requirements~\cite{holevo}.

In~\cite{frames-dc-bbm}, the authors introduced the
(ultimate) optimal protocol for transmitting orientations using a
quantum channel consisting of a system of spins. This protocol is
covariant and uses entanglement much in the same way as dense
coding~\cite{dense,ariano} does by requiring  Alice and Bob to
share the maximal entangled state of~\cite{ajv}  (see
Eq.~\ref{maxentangled} below). The results include the calculation
of the transmission error for large $N$ (or equivalently, the
error in the optimal estimation of a unitary transformation),
which shows an outstanding reduction as compared with the
previously known protocols. It should be emphasized however that
the improvement is achieved at the cost of keeping non-local
correlations between sender and recipient which, of course, is an
additional resource.

The aim of this paper is to  show that we can cut down the number
of spins to $N$  and still achieve a transmission error
asymptotically equal to that of the dense covariant coding
protocol.
%, without requiring Alice and Bob to share non-local correlations
We will show that this is not at odds with~\cite{frames-dc-bbm}
despite the apparent contradiction with the comments in the
previous paragraph. This economy of resources is mainly the result
of using efficiently the Hilbert space of the $N$ spins, which
span a number of equivalent irreducible representations of
$SU(2)$, as apposed to the protocol in~\cite{frames-bbm}. Note,
however, that the latter is the optimal one if an atom is used
instead of $N$ spins. We should also stress that the present
approach is entirely covariant. Thus, we resolve the
covariance-optimality puzzle discussed above.
%Though in some sense
%it was already resolved in~\cite{frames-dc-bbm}, one could quite
%fairly argue that this was not so because the use of non-local
%correlations between Alice and Bob in~\cite{frames-dc-bbm}
%invalidates the comparison with~\cite{frames-ps,
%frames-bbm,frames-ps2}, where no use was made of such resource .
%Both, the protocol we present in this paper and the non-covariant
%ones in~\cite{frames-ps2} are now on the same footing: they use
%the same number of spins and no shared entanglement. The former is
%covariant and optimal (at least asymptotically), as one would
%expect.
From the point of view of estimating $SU(2)$
transformations our results also mean an equally outstanding
reduction of the resources required to achieve an asymptotically
optimal estimation.

Let us assume that Alice has $N$ qubits at her disposal with which
she would like to either estimate an unknown $SU(2)$
transformation or communicate to Bob the orientation of a
trihedron. As mentioned above, the latter can in principle be
achieved regardless the existence of a shared reference frame if
we choose the $N$ qubits to be particles with spin. From now on we
will always refer to these $N$ qubits as spins for simplicity. The
most general preparation Alice can make is
\begin{equation}
|\Psi\rangle=\sum_{j,m,\alpha} \Psi^j_{m\alpha}|jm\alpha\rangle,
\qquad \sum_{j,m,\alpha} |\Psi_{m\alpha}^j|^2=1, \label{state Psi}
\end{equation}
where $j$ labels the irreducible representations of $SU(2)$ (i.e.,
$j(j+1)$ are the eigenvalues of ${\bf J}^2$, the total angular
momentum squared), $m$ are the $2j+1$ eigenvalues of $J_z$ ---which
label the elements of the standard orthonormal basis spanning the
$\bf j$ representation of $SU(N)$ of dimension $d_j=2j+1$--- and
$\al$ labels the $n_j$ different equivalent representations of
spin $j$ that show up in the Clebsch-Gordan series of $({\bf
1/2})^{\otimes N}$. One can compute $n_j$ to be
\begin{equation}
n_j={2j+1\over N/2+j+1}\pmatrix{N\cr N/2+j}  .
\label{n_j}
\end{equation}
We wish to view $\ket{\Psi}$ as a reference state to which Alice
will apply the unitary operation $U(g)=u^{\otimes N}(g)$;
$u(g)\in{\bf 1/2}$. Throughout this paper, $g$ will stand for
$SU(2)$ group parameters, such as the standard Euler angles
$g=(\alpha,\beta,\gamma)$. We will use the notation $gg'$ to
denote the parameters of the composition (product)
$U(g)U(g')\equiv U(gg')$, and $dg$ will stand for  the Haar
measure of $SU(2)$, which is left and right invariant under the
above composition, namely $d(gg')=d(g'g)=dg$, and normalized so
that $\int dg=1$.

If the operation (or one-qubit gate) $u(g)$ is unknown to Alice,
she can gain some knowledge about it by applying it to
$\ket{\Psi}$ to obtain a state $\ket{\Psi(g)}=U(g)\ket{\Psi}$ and
by performing an appropriate measurement over $\ket{\Psi(g)}$
afterwards. We will allow Alice to perform a completely general
positive operator valued  measure, or POVM, characterized by the
set of operators $\{O_r\}$, each one of them associated to a
possible outcome $r$. Alice can make a guess or have an estimate
of the parameter $g$ which will depend on the outcome she obtains.
Let us call $g_r$ the guess corresponding to outcome $r$. A
quantitative assessment of Alice's performance is given by the
averaged fidelity, defined as
\begin{equation}
\langle F\rangle =\sum_r\int dg\, F(g_r,g)\, p(r|g),
\end{equation}
where $F(g_r,g)\equiv |\tr[u^\dagger(g_r)u(g)]|^2/4$ is an
(squared) average over all input qubit $\ket{\phi}$ of how well
$u(g_r)\ket{\phi}$ compares to $u(g)\ket{\phi}$ ~\cite{ajv}, and
$p(r|g)$ is the probability of obtaining the outcome $r$ if the
unknown transformation is $u(g)$. In terms of group characters
$F(g_r,g)$ can also be written as
\begin{equation}
 F(g_r,g)=\frac{\chi_{1/2}^2(g_r^{-1}g)}{4}=\frac{1+\chi_1(g_r^{-1}g)}{4},
\end{equation}
 where
$\chi_j(g)$ is the character of the representation ${\bf j}$.
Quantum mechanics tell us that $p(r|g)=\tr[O_r \rho(g)]$, where
$\rho(g)=\ket{\Psi(g)}\bra{\Psi(g)}$. Note that we compute
$\langle F\rangle$ assuming that the {\em a priori} probability
for $u(g)$ is uniform with respect to the $SU(2)$ Haar measure.

Somewhat more speculatively, Alice could also use her $N$ spins to
transmit the orientation of an orthogonal trihedron, $\mathsf
n=\{\vec n^{(1)},\vec n^{(2)},\vec n^{(3)}\}$. In this case, she
would choose the state $\ket{\Psi}$ in such a way that the system
of spins had a physically observable magnitude that she could
correlate to $\mathsf n$~\cite{frames-lpt}(e.g., a magnetic or
electric quadrupole moment). She would then simply rotate the
system so that its orientation were that of $\mathsf n$ and would send
it to Bob. If we allowed him to perform a generalized measurement
$\{O_r\}$, he could infer from the outcomes the orientation of the
$N$-spin system and, hence, of Alice's trihedron $\mathsf n$.
Referred to an observer's reference frame $\mathsf n_0=\{\vec
x,\vec y,\vec z\}$, Alice's trihedron is $\mathsf n(g)=R(g)\mathsf
n_0$, where $R(g)$ is a rotation in 3-dimensional space. If $R(g)$ has
the unitary representation $u(g)$, the state Alice has prepared
and sent to Bob is again $\ket{\Psi(g)}$. Referred to the same frame,
the trihedron $\{\vec n^{(1)}_r, \vec n^{(2)}_r, \vec
n^{(3)}_r\}$ Bob guesses from the outcome $r$ of his measurement
should correspond to some $\mathsf n(g_r)=R(g_r) \mathsf n_0$
(Note that Bob does not know the actual value of $g_r$, since we
assume he does not know $\mathsf n_0$).
The quality of the transmission can, thus, be quantified through
the averaged Holevo's error\cite{holevo, frames-bbm}
\begin{equation}
\langle h\rangle = \sum_r\int dg\,h(g_r,g) p(r|g),
\end{equation}
where $h(g_r,g)=\sum_{a=1}^3|\vec n^{(a)}(g_r)-\vec
n^{(a)}(g)|^2=6-\chi_1(g_r^{-1}g)$. This shows that the two
problems we are dealing with, i.e., estimation of $SU(2)$
transformations and transmission of frames/trihedra, are formally
the same. Throughout the rest of the paper, we will concentrate
in~$\langle\chi_1\rangle$
\begin{equation}
\langle\chi_1\rangle=\sum_r\int dg\, \chi_1(g_r^{-1} g)
\tr\left[O_r\rho(g)\right],
\end{equation}
from which we straightforwardly obtain either $\langle
F\rangle=(1+\langle\chi_1\rangle)/4$ or $\langle
h\rangle=6-\langle\chi_1\rangle$, depending on the problem we are
interested in.
Our conclusions directly apply to the two problems above, which we
may simply regard as two different aspects of the same topic.

As mentioned in the introductory comments, the optimal scheme (the
one that leads to the maximal $\langle\chi_1\rangle$) requires
$\ket{\Psi}$ to be the maximally entangled $2N$-spin state
\begin{equation}\label{maxentangled}
\ket{\Phi}=\sum_j a_j\ket{\Phi^j}\equiv\sum_j{a_j\over
\sqrt{d_j}}\sum_{m=-j}^j\ket{jm}_A\ket{jm}_B , \label{state Phi}
\end{equation}
where $j$ runs from the highest total spin $J\equiv N/2$ to $1/2$
($0$) for $N$  odd (even), and the action of $SU(2)$ to be
\begin{equation}
U(g)=U_A(g)\otimes \id_B=[u(g)]^{\otimes N}_A\otimes\id_B ,
\label{action of SU(2)}
\end{equation}
where $A$  refers to the first $N$ (active) spins and $B$ to the
other $N$ (spectator) spins (in the  dense covariant coding
approach of~\cite{frames-dc-bbm}, $A$ and $B$ refer to Alice and
Bob respectively). Within this framework we obtain for large $N$
\begin{equation}\label{chi-entangled}
\langle\chi_1^{\rm entgl}\rangle=3-{4\pi^2\over N^2}
+{24\pi^2\over N^3}+\dots .
\end{equation}

We now realize that we can make do with just $N$ spins if we
replace the $d_j$ degrees of freedom involved in each one of the
$\ket{jm}_B$ by those corresponding to the $n_j$ equivalent
representations ${\bf j}$ in~(\ref{state Psi}). More precisely, we
assign to each $m$ a unique $\alpha$ (see Eq.~\ref{state Psi}),
which we denote by $\alpha_m$, and entangle these two degrees of
freedom. Clearly, the quantum correlations of~(\ref{state Phi})
are exactly those of  $\ket{\Psi}=\sum_j\ket{\Psi^j}$, where
\begin{equation}
\ket{\Psi^j}={1\over\sqrt{d_j}}\sum_{m=-j}^j\ket{jm\alpha_m}.
\label{state Psi^j}
\end{equation}
It is important to note that this entanglement of degrees of
freedom can be established in any of the $\bf j$ invariant
subspaces but in the $\bf J$ subspace (the one corresponding to
the highest spin, $N/2$), since~(\ref{n_j}) implies
\begin{equation}
n_j\ge d_j \qquad \mbox{if}\; j<J; \qquad n_{J}=1.
\end{equation}
Hence $\ket{\Phi^j}$ and $\ket{\Psi^j}$ have the same
entanglement for $j<J$, whereas $\ket{\Psi^{J}}=\sum_m
\Psi^{J}_m\ket{Jm}$ has no entanglement at all (the ${\bf J}$
representation occurs only once in the Clebsch-Gordan series of
${\bf 1/2}^{\otimes N}$). It is also important to note that the
index $\alpha$ that labels the equivalent representations does not
transform under $SU(2)$. Hence, the action of this group over
$\ket{\Psi}$ is still given by~(\ref{action of SU(2)}), where now
$B$ refers to the `$\alpha$ degrees of freedom'.

We would like to stress that the $n_j-d_j$ equivalent
representations that do not show up in~(\ref{state Psi^j}) are
actually sterile. They cannot be used for the problems at hand, as
shown by the following argument. The action of~(\ref{action of
SU(2)}) on a general state belonging to the direct sum of all the
equivalent representations ${\bf j}$  yields
$\ket{w(u)}=\sum_{m\alpha}w_{m\alpha} \,u^{\otimes
N}\ket{jm\alpha}$. Let
$\ket{\phi}=\sum_{m'\alpha'}\phi_{m'\alpha'}\ket{jm'\alpha'}$ be
another state belonging to the same subspace. We have
$\langle\phi|w(u)\rangle=\sum_{mm'}(\sum_\alpha\phi^*_{m'\alpha}w_{m\alpha})\D
jm{m'}(u)$, where $\D jm{m'}$ is the standard $d_j$-dimensional
unitary matrix representation of $SU(2)$. We can find at least
$n_j-d_j$ ($n_j$-dimensional) `vectors'
$(\eta_{a1},\eta_{a2},\eta_{a3},\dots)$, $a=1,2,\dots, n_j-d_j$
orthogonal to all the $d_j$ `vectors'
$(w_{m1},w_{m2},w_{m3},\dots)$, $m=-j,-j+1,\dots,j$. Defining
$\phi_{m\alpha}=\varphi_{pm}\eta_{a\alpha}$, where the complex
numbers $\varphi_{pm}$, $p=1,2,\dots,d_j$, are chosen so that
$\sum_m\varphi^*_{pm}\varphi_{qm}=\delta_{pq}$,  we see that the
orthogonal complement of $\{\ket{w(u)}\}_{u\in SU(2)}$ has at
least dimension $d_j(n_j-d_j)$, since $\langle\phi|w(u)\rangle=0$
for all $u\in SU(2)$. Hence, the signal state can  span at most
$d_j$ ${\bf j}$-invariant subspaces.

Keeping all the above in mind and recalling from\cite{frames-bbm}
that $\ket{JJ}$ is optimal when only one of the equivalent
representations $\bf j$ is allowed, it is tempting to state that
\begin{equation}
\ket{\Psi_{\{a\}}}=a_J \ket{JJ}+\sum_{j<J}
{a_j\over\sqrt{d_j}}\sum_{m=-j}^j\ket{jm\alpha_m} \label{new Psi}
\end{equation}
is optimal  for both the estimation of $SU(2)$ transformations and
the transmission of frames (for a suitable  set of real
coefficients $\{a\}$ obeying the normalization condition $\sum_j
|a_j|^2=1$). This is not entirely (but almost) right  because of
the small asymmetry introduced by the highest spin component
$\ket{\Psi^J}$. However, we will show below that the maximal
$\langle\chi_1\rangle$ we can obtain using~(\ref{new Psi}) differs
from the optimal one~\cite{frames-dc-bbm} only by terms that
vanish asymptotically as $1/N^3$. This means that $N$ spins
suffice to asymptotically attain the dense covariant coding bound,
which uses an entangled state of $2N$ spins.

We first show that a continuous rank one POVM does exist for signal
states of the type~(\ref{new Psi}). Using Schur's lemma, one can
readily see that
\begin{equation}
\int dg\; U(g)\, O^j\, U^\dagger(g)
={\id^j_A\otimes \tr_A O^j\over d_j}
\label{property}
\end{equation}
over each irreducible subspace of $SU(2)$ of dimension $d_j$. Here
$\tr_A$ is the partial trace over subsystem $A$ (the `$m$ degrees
of freedom'). If $O^j$ is rank one, we have
$O^j=d_j^2\ket{\phi^j}\bra{\phi^j}$ and condition~(\ref{property})
is $\tr_A(\ket{\phi^j}\bra{\phi^j})=\id^j_B/d_j$, which implies
that $\ket{\phi^j}$ is a maximally entangled state over each
irreducible representation (except ${\bf J}$). We may chose it to
be of the form~(\ref{new Psi}) without any loss of generality.
Hence, the continuous POVM is
\begin{equation}
O(g)=U(g)\ket{\Psi_{\{b\}}}\bra{\Psi_{\{b\}}}U^\dagger(g),
\end{equation}
where $\ket{\Psi_{\{b\}}}$ is defined as in~(\ref{new Psi}), and
the set $\{b\}$ is given by  $b_j=d_j$ for  $j<J$,
$b_J=\sqrt{d_J}$. POVMs with a finite number of outcomes can also
be found following~\cite{frames-bbm}.

We are now in the position to compute $\langle\chi_1\rangle$ for
the signal states~(\ref{new Psi}). This will provide  a lower
bound for $\langle\chi_1^{\rm opt}\rangle$, the averaged $\chi_1$
of the optimal $N$-spin scheme. Recalling the invariance of $dg$
and Schur's lemma one gets
\begin{eqnarray}
\langle\chi_1\rangle&=&\int dg\int dg'\, \chi_1(g^{-1}g')
 \,\tr[O(g) \rho(g')]\nonumber\\
&=&
\int dg\int dg' \,\chi_1(g^{-1}g') \,
\tr[\ket{\Psi_{\{b\}}}\bra{\Psi_{\{b\}}} \rho(g^{-1}g')]
\nonumber\\
&=& \int dg\, \chi_1(g)|\bra{\Psi_{\{a\}}}U(g)\ket{\Psi_{\{b\}}}|^2
\nonumber\\
&=&{1\over 3}\sum_{jl}a^ja^l\left[b^jb^l\,\tr_{\bf 1}\left(
\rho^j\otimes\tilde\rho^l \right)\right],
\end{eqnarray}
where we have defined the operators $\rho^j$ and $\tilde\rho^l$
through the relations
$\sum_{j}a^jb^j\rho^j=\tr_B(\ket{\Psi_{\{b\}}}\bra{\Psi_{\{a\}}})$
and  $\sum_{l}a^l b^l\tilde\rho^l=\tr_B(\ket{\tilde\Psi_{\{b\}}}
\bra{\tilde\Psi_{\{a\}}})$. The state $\ket{\tilde\Psi_{\{a\}}}$
is the transformed of $\ket{\Psi_{\{a\}}}$ under time reversal and
$\tr_{\bf 1}$ is the trace over the representation ${\bf 1}$
invariant subspace, i.e., $\tr_{\bf 1}
O=\sum_{m=-1}^1\bra{1m}O\ket{1m}$. For $j<J$ we see that
$\rho^j=\tilde\rho^j=\id^j/d_j$, whereas $\rho^J=\ket{JJ}\bra{JJ}$
and $\tilde\rho^J=\ket{J-J}\bra{J-J}$. Using that
\begin{equation}
\tr_{\bf 1}\left(\ket{jm}\bra{jm'}\otimes\id^l\right)={3\delta_{mm'}\over d_j}
\end{equation}
for $j+l\ge1\ge|j-l|$ (it vanishes otherwise), along with
 $d_J\langle JJ;J-J|10\rangle^2=3J/(J+1)$, we obtain
\begin{equation}
\langle\chi_1\rangle=1+{\mathsf a}^t{\mathsf M}{\mathsf a} ,
\end{equation}
where ${\mathsf a}^t=(a^{J},a^{J-1},\dots)$ is the transpose of
$\mathsf a$, and $\mathsf M$ is the $n\times n$ tridiagonal matrix
\def\xxx{\phantom{\displaystyle{.\over .}}}
\def\zzz{\hspace{.5em} }
\begin{equation}\label{matrix}
\mathsf{M}=\pmatrix{
 {-1\over J+1}     & {1\over \sqrt d_J}              &         &   &  &  &\xxx\cr
 {1\over \sqrt d_J}& 0                               &   1     &   & & \mbox{\LARGE0} &\xxx\cr
                   & 1 & 0 &\zzz 1                   &         &   &\xxx\cr
                   &   &\raisebox{1.3ex}[0ex][0ex]{1}&  \ddots & \ddots & &\xxx\cr
                   &   &                             & \ddots  &       & 1&\xxx\cr
 \hspace{.5cm} \raisebox{2.0ex}[1.5ex][0ex] {\LARGE0}\hspace{-.5cm} & & & &  1 \zzz & 0 &\zzz 1\xxx\cr
                   &   &   &  &  &    1  &   \zzz      0\xxx } \ ,
\end{equation}
where here and throughout the rest of the paper we will only
consider $N$ odd ($J$ half integer, $n=J+1/2$) for simplicity. The
maximum value of $\langle\chi_1\rangle$ is
\begin{equation}
\langle\chi_1\rangle=1-2\lambda_0,
 \label{la_0}
\end{equation}
where $-2\lambda_0$ is the largest eigenvalue of $\mathsf M$. This
can be computed easily by noticing that the characteristic
polynomial of $\mathsf M$, defined to be
$P^{J}_n(\lambda)=\det({\mathsf M}+2\lambda\,\id)$, satisfies the
recursive relation of the Tchebychev
polynomials~\cite{abramowitz}, namely: $P^{J}_n(\lambda)=2\la
P^{J}_{n-1}(\la)-P^{J}_{n-2}(\la)$. Hence, $P^{J}_n(\la)$ is a
linear combination of them. One can easily check that the explicit
solution is
\begin{equation}
P_n(\la)=U_n(\la)-{2\over2n+1}
U_{n-1}(\la)+{2n-1\over2n}U_{n-2}(\la),
\end{equation}
where we have defined $P_n(\lambda)\equiv P^{n-1/2}_n(\la)$ and
$U_n(\cos\theta)=\sin[(n+1)\theta]/\sin\theta$ are the Tchebychev
polynomials of the second kind. Hence, the smallest zero of
$P_n(\la)$, which is $\la_0\equiv\cos\theta_0$ in~(\ref{la_0}),
can be easily computed in the large $n$ limit expanding around
$\lambda_0=-1$, i.e, $\theta_0=\pi(1-n^{-1}+a n^{-2}+b
n^{-3}+\dots)$. We find
\begin{equation}
\langle\chi_1\rangle=3-{4\pi^2\over N^2}+{8\pi^2\over N^3}+\dots .
\end{equation}
Recalling that $\langle\chi_1\rangle\le\langle\chi_1^{\rm
opt}\rangle\le\langle\chi_1^{\rm entgl}\rangle$
and~(\ref{chi-entangled}), we conclude that
\begin{equation}
\langle\chi_1^{\rm opt}\rangle=3-{4\pi^2\over N^2}+O(1/N^3) .
\end{equation}
Therefore, the theoretical limit imposed by Eq.~\ref{chi-entangled}
is asymptotically reached with just half the number of spins and
without non-local correlations being shared between Alice and Bob.

In summary. The  internal ``$\alpha$" degrees of freedom
associated to repeated irreducible representations in the
Clebsch-Gordan series of ${\bf (1/2)}^{\otimes N}$ do not
transform under  the $SU(2)$ group or rotations and, therefore,
they are not directly useful to encode such transformations.
However, they can be entangled to proper $SU(2)$ degrees of
freedom, such as $m$, to yield an outstanding improvement over
some previously known estimation and communication  protocols.
This entanglement is formally the same as that of dense covariant
coding but it is achieved with half the number of spins and
without preestablished quantum correlations between distant
parties.  Furthermore, since these degrees of freedom are
invariant under rotations, the resulting protocol is manifestly
covariant.

 We acknowledge financial support from
Spa\-nish Ministry of Science and Technology project
BFM2002-02588, CIRIT project SGR-00185, and QUPRODIS working group
EEC contract IST-2001-38877. RMT acknowledges the hospitality of
the Quantum Information Theory group at Pavia.

\end{document}